# The immediate effect of kangaroo mother care on Mother-infant inter-brain synchrony and infant brain function

Yu Liu[1#], Jiayang Xu[2#], Tianzi Wang[1#], Zichen Shi[2], Xiang Chen[1], Yanting Kong[1], Lianli Chen[1], Sha Sha[1], Shanbao Tong[2], Chuhan Dong[1], Guanghai Wang[3,4,5,6*], Xiaoli Guo[2,*], Fei Bei[1,*]

[1]Department of Neonatology, National Children's Medical Center, Shanghai Children's Medical Center, Affiliated to School of Medicine Shanghai Jiao Tong University, Shanghai, China.

[2]School of Biomedical Engineering, Shanghai Jiao Tong University, Shanghai, China.

[3]Department of Developmental and Behavioral Pediatrics, Shanghai Children's Medical Center, School of Medicine, Shanghai Jiao Tong University, Shanghai, China.

[4]Pediatric Translational Medicine Institute, Shanghai Children's Medical Center, School of Medicine, Shanghai Jiao Tong University, Shanghai, China.

[5]Shanghai Center for Brain Science and Brain-Inspired Technology, Shanghai, China.

[6]Ministry of Education–Shanghai Key Laboratory of Children's Environmental Health, Xinhua Hospital, School of Medicine, Shanghai Jiao Tong University, Shanghai, China.

[*]**Corresponding authors:**

Dr. Fei Bei at Department of Neonatology, National Children's Medical Center, Shanghai Children's Medical Center, Affiliated to School of Medicine Shanghai Jiao Tong University, Shanghai, China, e-mail: beifei@scmc.com.cn.

Dr. Xiaoli Guo at School of Biomedical Engineering, Shanghai Jiao Tong University, Shanghai, China, e-mail: meagle@sjtu.edu.cn.

Dr. Guanghai Wang at Department of Developmental and Behavioral Pediatrics, Shanghai Children's Medical Center, School of Medicine, Shanghai Jiao Tong University, Shanghai, China; Pediatric Translational Medicine Institute, Shanghai Children's Medical Center, School of Medicine, Shanghai Jiao Tong University, Shanghai, China; Shanghai Center for Brain Science and Brain-Inspired Technology, Shanghai, China; Ministry of Education–Shanghai Key Laboratory of Children's

Environmental Health, Xinhua Hospital, School of Medicine, Shanghai Jiao Tong University, Shanghai, China, e-mail: wang-guanghai@163.com.

#Yu Liu, Jiayang Xu and Tianzi Wang contributed equally to this work.

## Abstract

Kangaroo mother care (KMC) is an intervention involving skin-to-skin contact that promotes physiological stability and supports long-term neurodevelopment in preterm infants. However, the underlying neurophysiological mechanisms remain unclear. We aimed to investigate the immediate effects of the first KMC on infants' brain function, mother-infant inter-brain synchrony, as well as their associations. Fifty-eight preterm infants (gestational age < 32 weeks or birth weight < 1500 g) and their mothers underwent synchronous dual-electroencephalography recording before and during the first KMC session. Infant brain function was assessed via power spectrum energy and graph theory-based network metrics, and mother-infant inter-brain synchrony was quantified using phase-locking value (PLV), from which inter-brain density and inter-brain strength were calculated. Correlation analyses were performed between infant intra-brain metrics and inter-brain synchrony indicators.During the first KMC, preterm infants showed enhanced theta, alpha, and beta power alongside reduced relative delta power, while brain network topological metrics remained stable. Concurrently, mother-infant inter-brain synchrony was significantly enhanced across all frequency bands, as evidenced by increased inter-brain density and strength (all $p < .001$). Furthermore, in the alpha band, inter-brain strength correlated positively with infant local efficiency and clustering coefficient, and in the beta band, it was positively correlated with infant small-worldness. The first KMC session can immediately enhance both preterm infant single-brain activity and mother-infant inter-brain synchrony. The strength of inter-brain synchrony is associated with the infant's intra-brain network organization, suggesting that KMC may promote intra-brain development in preterm infants via enhancing mother-infant inter-brain synchrony.

# Introduction

High-quality early parent-infant interaction is crucial for children's social, emotional, and language development, and has been associated with lower rate of autism spectrum disorders (ASD) and attention deficit hyperactivity disorder (ADHD) [1-4]. However, for preterm infants, during their early stay in the Neonatal Intensive Care Unit (NICU), they are not only exposed to stressful stimuli such as noise, light, and procedural pain, but also face early separation from their mothers, depriving them of essential parent-infant interaction. Kangaroo Mother Care (KMC) is a parent-infant skin-to-skin care practice for preterm infants, recommended by the World Health Organization as a core intervention to improve outcomes for preterm infants. Current guidelines indicate that KMC not only stabilizes neonatal clinical status and shortens hospital stay but also exerts lasting beneficial effects on long-term cognitive development, social function, and emotional interaction, with earlier initiation being more advantageous[5-8]. Yet, the neural mechanism underlying KMC remained little explored.

Current neurophysiological evidence has begun to elucidate the impact of KMC on the infants' brain activity. For example, using a heel lance paradigm combined with EEG, Jones et al. [9] found that parental skin-to-skin contact modulates the infants' cortical response to noxious stimuli and enhances subsequent cortical processing. Similarly, Kaffash et al. [10] demonstrated that, at the same post-menstrual age, preterm infants who received KMC exhibited higher EEG complexity than non-KMC infants, indicating more advanced brain network development.

Although these studies provide valuable insights into infant neurodevelopment, they primarily focus on the infant's brain activity and consider the mother only as part of the environmental context rather than an active participant in dyadic interaction. Given that high-quality parent-infant interaction is inherently a bidirectional, dynamic process of physiological and neural coordination[11-13], we hypothesize that the neurophysiological mechanism of KMC involves not only the optimization of the infant's intra-brain activity but also the emergency of mother-infant inter-brain synchrony, with these intra- and inter-brain processes forming a reciprocally interactive system.

Hyperscanning provides a robust method for investigating dyadic brain interactions, with studies in 2- to 6-month-olds confirming that mother-infant intimate contact, gaze, and vocalizations enhance inter-brain synchrony and physiological signal synchrony [14]. Existing research suggests that KMC can promote long-term behavioral synchrony between parents and preterm children, achieving levels comparable to those observed in full-term infants during adulthood. However, parent-infant behavioral synchrony is not significantly evident in early infancy, limiting its utility for early prediction and real-time assessment[15]. Behavioral synchrony serves as an external indicator of underlying physiological and neural synchrony. We thus hypothesize that even in the absence of observable behavioral synchrony, close physical contact may foster inter-brain synchrony in early stages. The intrinsic relationship between mother-infant inter-brain synchrony and preterm infants' individual neurodevelopment remains incompletely characterized, particularly whether initial contact during the first KMC session induces immediate inter-brain synchrony or whether such synchrony correlates with intra-brain functions.

To answer this question, the present study employed hyperscanning technology to simultaneously record EEG from preterm infants and their mothers during the first KMC session, as well as during a 5-10 minutes period before its initiation. By concurrent intra-brain and inter-brain analyses, we aim to unravel the neurophysiological mechanisms underlying the immediate effects of KMC on infants' brain activity, mother-infant inter-brain synchrony as well as intra-inter-brain associations.

## Methods

### Participants

The present study was a secondary/subgroup analysis of participants allocated to the KMC intervention arm of a larger randomized controlled trial (ChiCTR2400088871). The inclusion criteria were: 1) singleton preterm infants admitted to the NICU of Shanghai Children's Medical Center within 24 hours after birth between April 1, 2024 and December 31, 2025, with the gestational age< 32 weeks or the birth weight <1500 g; 2) infants who received KMC. We excluded infants with major congenital malformations, severe asphyxia, grade III or higher intraventricular hemorrhage, seizures. A total of 60 preterm infants along with their mothers were initially enrolled. Among those two mother-infant dyads were excluded due to poor EEG data quality that precluded analysis, resulting in a final sample of 58 mother-infant dyads (n = 58) **(Fig.1)**. Ethical approval was obtained from the institutional review board of Shanghai Children's Medical Center (SCMCIRB-K2024049-1). Written informed consent was obtained from all the participants (parents on behalf of their children).

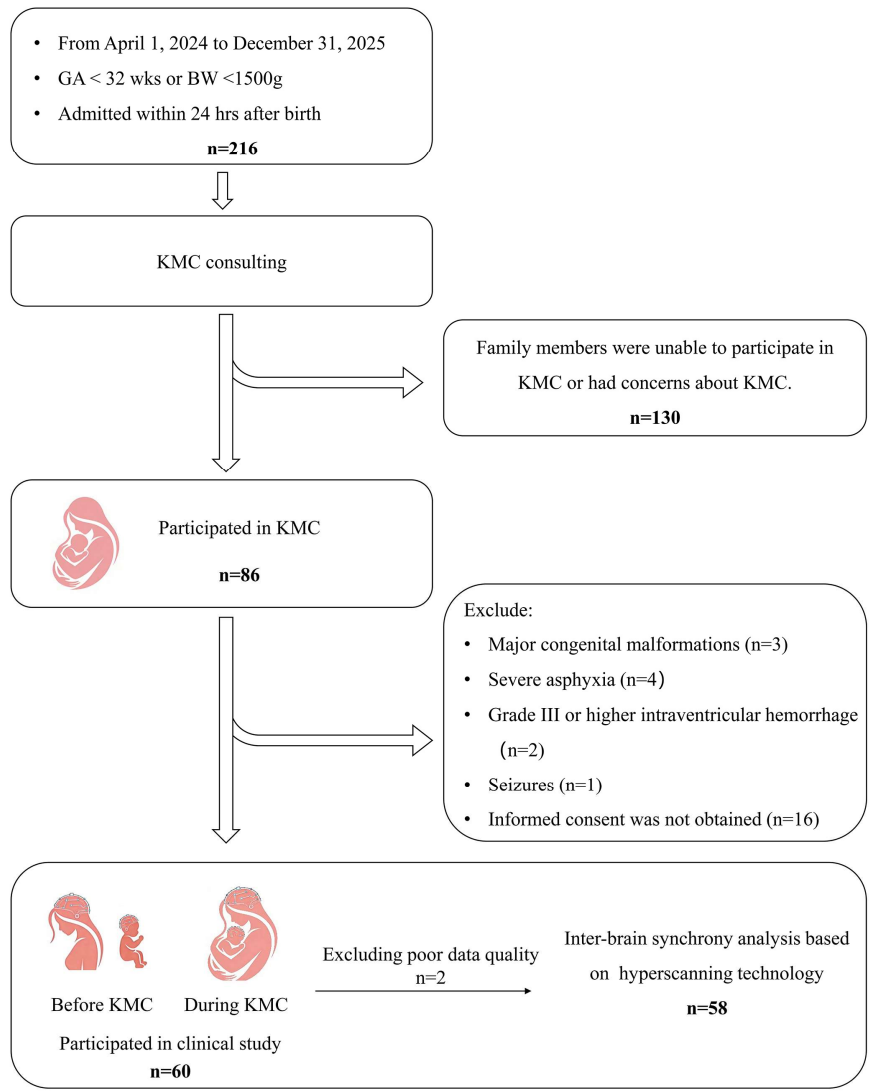

**Fig1. Flowchart of participant enrollment and analysis.**

## KMC paradigm

KMC was conducted in a quiet, dimly lit single room with the ambient temperature maintained between 24°C and 26°C. To standardize conditions, infants were fed and diapered 30 minutes prior to the KMC session, after which no further procedures were performed. Mothers were positioned in a semi-recumbent posture on a dedicated KMC reclining chair, with the backrest inclined at approximately 60° to ensure their comfort. Before KMC, the infants remained in the incubator while the mothers sat quietly in the chair for 5-10 minutes. During KMC, the infant was placed in a prone position with the head turned to one side on the mother's bare chest, positioned between her breasts. A trained nurse guided the mother to support the infant by placing one hand on the infant's back and the other under the buttocks to prevent slipping. A blanket was used to cover the infant's back and body. The duration of KMC was 30 minutes. To minimize artifacts in the EEG recording, mothers were instructed to avoid strenuous physical activity and significant emotional fluctuations during the whole process. The entire procedure was conducted under the supervision of professional pediatricians and nursing staff. KMC was immediately discontinued in cases of hypothermia, unstable vital signs, vomiting, or inconsolable agitation.

## Dual-EEG data acquisition

Specially trained nurses or physicians proficient in EEG procedures applied EEG caps to both mothers and infants according to standardized protocols. Electrodes were positioned according to the international 10–20 system, with FCz as the reference electrode and AFz as the ground electrode. All electrode impedances were maintained below 10 kΩ, and the sampling rate was set at 1000 Hz. The EEG acquisition system consisted of one 32-channel EEG wireless system for mothers and one 10-channel EEG wireless system for infants (NeuSen.W32, Neuracle, China). Synchronization triggers were transmitted to both synchronization boxes before each session to ensure temporal alignment between the two EEG systems. Simultaneous EEG was recorded from both mother and infant throughout the entire KMC session. Two panoramic cameras were used to monitor the entire procedure, recording gaze direction, facial expressions, and limb movements. Video data were transmitted in real time to the EEG acquisition interface, enabling synchronized recording of EEG and behavioral signals at 25 frames per second.

## EEG Preprocessing

EEG preprocessing was conducted in MATLAB (R2019a, The MathWorks Inc., Natick, MA, USA) utilizing the EEGLAB toolbox. The procedure included band-pass filtering, removal of ocular and electromyographic artifacts via independent component analysis (ICA), and exclusion of segments with amplitudes exceeding ±100 µV. For infants, data from 9 EEG channels (Fp1, Fp2, Cz, C3, C4, T3, T4, O1, O2) were analysis. For mothers, peripheral temporal channels (T7, T8, Tp9, Tp10) were excluded due to significant muscular artifact contamination, leaving 26 channels (Fp1, Fp2, F3, F4, Fz, F7, F8, FC1, FC2, FC5, FC6, Cz, C3, C4, CP1, CP2, CP5, CP6, Pz, P3, P4, P7, P8, O1, O2, Oz) for analysis. The preprocessed continuous EEG signals were band-pass filtered to extract frequency-specific signals. The frequency bands of interest in this study were delta (0.5–3 Hz), theta (3–6 Hz), alpha (6–9 Hz), and beta (9–12 Hz), corresponding to infant-specific frequency bands. Finally, EEG data before and during KMC were segmented into 2-second epochs for further analysis.

## Intra-brain Analysis of infants' EEG c

To analyze changes in infants' brain function, spectral power and graph-theory-based brain network measures were calculated. Spectral power analysis was conducted using the welch method, a widely used nonparametric approach for power spectral density (PSD) estimation. Absolute and relative spectral power were calculated for each frequency band of interest. Continuous EEG signals were divided into overlapping segments, each multiplied by a window function such as Hanning window to minimize spectral leakage. The Fourier transform was computed for each segment to obtain its PSD, and the averaged PSD within each frequency band across all the segments was used as the absolute spectral power. Relative spectral power was then computed by normalizing the absolute power within each frequency band to the total power across all analyzed frequency bands.

Furthermore, functional brain networks were constructed using the phase-locking value (PLV) algorithm, which is one of the most commonly used methods for estimating undirected functional connectivity[16]. A PLV value of 1 indicates perfect phase coupling between two signals, whereas a PLV value of 0 indicates completely asynchronous phase fluctuations. For the constructed functional brain networks, the top 25% of connection strengths were retained to generate sparse adjacency matrices. The following graph-theory-based measures were then calculated: (1) global efficiency, (2) local efficiency, (3) clustering coefficient, (4) small-worldness.

### Inter-brain Analysis of mother-infant EEG

For the segmented narrow-band signals of each frequency band, inter-brain synchrony was also quantified using the PLV. To eliminate spurious connections arising from shared stimuli or non-interaction-related similarities, a surrogate-based thresholding procedure was applied. Specifically, within-dyad trials were randomly shuffled 1000 times to generate surrogate inter-brain connectivity matrices. Paired $t$-tests were then performed between the real PLV matrix and the surrogate matrices, and only statistically significant edges ($p < .05$) were retained. Based on the inter-brain matrix after threshold, two metrics were further extracted: inter-brain density (IBD, defined as the proportion of retained edges relative to the all edges) and inter-brain strength (IBS, defined as the mean strength of the retained inter-brain connections).

### Statistical Analysis

Statistical analyses were performed using Jamovi 2.3.21. Continuous variables were tested for normality. Normally distributed data are presented as mean ± standard deviation (SD), whereas non-normally distributed data are reported as median (P25, P75). Categorical variables are presented as frequencies (percentages). Paired t-test assessed differences in mother–infant inter-brain network density and strength before and during KMC as well as changes in preterm infants' network metrics and spectral power indices. The Pearson's $r$, or Spearman's $\rho$ for non-normally distributed variables, was used to examine relationship between intra-brain and inter-brain measurements. A *p*-value < .05 was considered statistically significant.

## Results

### Clinical characteristics of mother-infant dyads

The baseline characteristics of the study population are summarized in **Table 1**. The study included 58 mother-infant dyads. This includes 32 (32/58, 55.2%) males and 26 (26/58, 44.8%) females. The average gestational age is 29.8±1.7 weeks, the average birth weight is 1228.5 ± 250.8g. The first KMC occurred at a postmenstrual age of 33.7 (32.5, 34.4) weeks. At the time of KMC, infants had an average body weight of 1500.4 ± 173.7 g and a median postmenstrual age of 33.7 (32.5, 34.4) weeks. For mothers, the mean pregnancy age is 33.5±4.4 years. Cesarean delivery rate is 79.3% (46/58). Maternal complications included 22 (22/58, 37.9%) gestational diabetes cases, 13 (13/58, 22.4%) gestational hypertension cases

**Table 1. Baseline characteristics of the study population.**

| Characteristic | Value |
| --- | --- |
| **Infant General Characteristics** | |
| Male sex (n, %) | 32 (55.2%) |
| Gestational age (weeks) | 29.8 ± 1.7 |
| Birth weight (g) | 1228.5 ± 250.8 |
| 1-minute Apgar score | 9 (7,10) |
| 5-minute Apgar score | 10 (9,10) |
| **Maternal Characteristics** | |
| Maternal age (years) | 33.5 ± 4.4 |
| Cesarean section (n, %) | 46 (79.3%) |
| Gestational diabetes (n, %) | 22 (37.9%) |
| Gestational hypertension (n, %) | 13 (22.4%) |
| Premature rupture of membranes ≥ 18h (n, %) | 14 (24.1%) |
| **Infant-Related Outcomes** | |
| ROP requiring surgery (n, %) | 3 (5.2%) |
| Severe BPD (n, %) | 6 (10.3%) |
| NEC stage ≥ IIA (n, %) | 5 (8.6%) |
| PVL (n, %) | 6 (10.3%) |
| Early-onset sepsis (n, %) | 20 (34.5%) |
| Late-onset sepsis (n, %) | 4 (6.9%) |
| Surgical treatment (n, %) | 4 (6,9%) |
| Death (n, %) | 0 (0%) |
| **Infant Status at KMC** | |
| Weight (g) | 1500.3 ± 173.7 |
| Postnatal age (days) | 24.0 (16.0, 32.5) |
| Postmenstrual Age (weeks) | 33.7 (32.7, 34.4) |

Values are presented as mean ± standard deviation (SD) or median (interquartile range, IQR) for continuous variables and number (percentage) for categorical variables.

ROP: retinopathy of prematurity; BPD: bronchopulmonary dysplasia; NEC: necrotizing enterocolitis; PVL: periventricular leukomalacia; KMC: kangaroo mother care.

## EEG power and brain network metrics of preterm infants during first KMC

Significant alterations in infant single-brain EEG metrics during KMC were primarily evident in spectral power **(Fig. 2A-B)**. In comparison to the pre-KMC period, absolute power during KMC increased significantly in almost all bands, including delta (189.35 ± 12.851 vs. 212.15 ± 10.636 µV², $t(57) = -1.74$, $p = .087$, Cohen's d = -0.229), theta (7.99 ± 0.569 vs. 10.26 ± 0.857 µV², $t(57) = -2.48$, $p = .016$, Cohen's d = -0.326), alpha (3.16 ± 0.248 vs. 4.23 ± 0.361 µV², $t(57) = -2.67$, $p = .01$, Cohen's d = -0.351), and beta (2.08 ± 0.159 vs. 2.73 ± 0.183 µV², $t(57) = -3.27$, $p = .002$, Cohen's d = -0.43) bands(Fig. 2A). However, for the relative power, a significant decrease was observed in the delta band during KMC (0.931 ± 0.003 vs.0.923 ± 0.004, $t(57) = 2.07$, $p = .043$, Cohen's d = 0.271), whereas relative power increased significantly in the alpha (0.017 ± 0.002 vs. 0.019 ± 0.001, $t(57) = -2.18$, $p = .033$, Cohen's d = -0.287) and beta (0.011 ± 0.001 vs.

0.012 ± 0.001, *t* (57) = -2.15, *p* = .036, Cohen's d = -0.282) bands. No significant difference was found for relative power in the theta band (*p* > .05) **(Fig. 2B).**

Comparing of functional network topology metrics across frequency bands revealed no statistically significant differences between before KMC and during-KMC periods for global efficiency, local efficiency, clustering coefficient, or small-worldness (all *p* > .05, **Fig. 2C–F**).

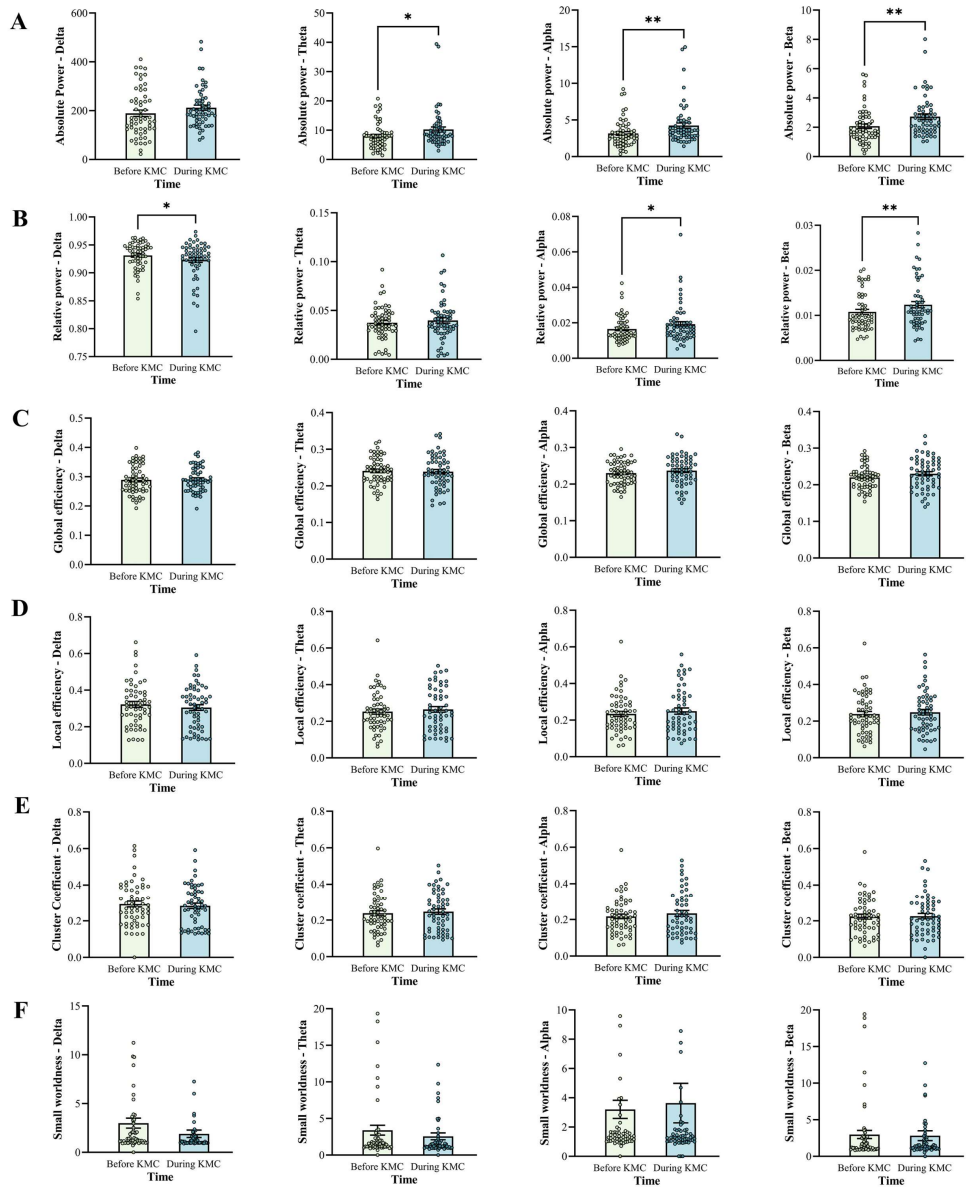

**Fig 2. Infant EEG power and functional network topology during KMC.** Absolute power (A) and relative power (B) in the delta, theta, alpha, and beta bands before and during KMC. Network

topological properties derived from infant EEG functional connectivity in the delta, theta, alpha, and beta bands, including global efficiency (C), local efficiency (D), clustering coefficient (E), and small-worldness (F), shown for the before KMC and during KMC conditions. Dots denote individual infants, bars represent group means, and error bars indicate SEM. *$P < .05$; **$P < .01$.

## Mother–infant inter-brain synchrony during first KMC

Mother-infant inter-brain functional connectivity was significantly enhanced during KMC **(Fig. 3)**. Connectivity matrices illustrated a marked increase in inter-brain synchronous connections between mothers and preterm infants during KMC compared to the before KMC period, with broader distributions of connections observed across the delta, theta, alpha, and beta bands **(Fig. 3A, C, E, G)**.

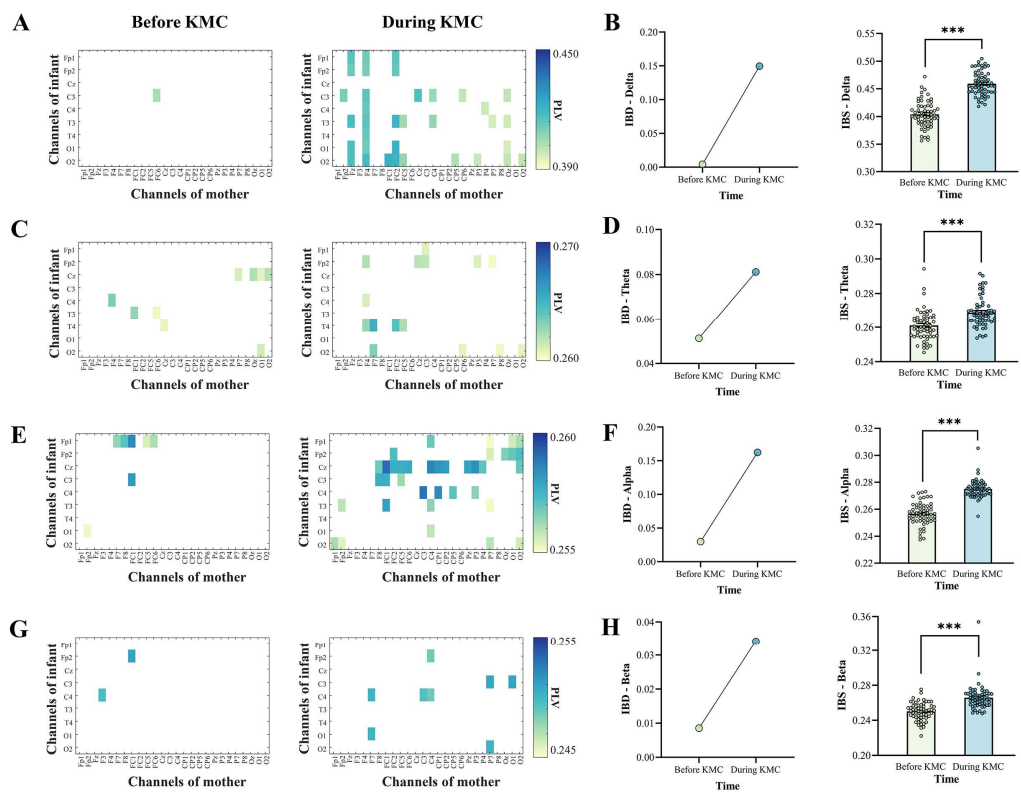

**Fig 3. Mother–infant inter-brain connectivity during kangaroo mother care.** Interbrain connectivity matrix (A, C, E, G) show PLV-based inter-brain connectivity matrices between maternal and infant EEG channels in the delta (A), theta (C), alpha (E), and beta (G) frequency bands, before and during KMC. The color scale represents the magnitude of PLV values, with warmer colors indicating stronger synchronization. Line plots (B, D, F, H, left) illustrate changes in inter-brain density (IBD). Bar plots (right) show inter-brain strength (IBS) before and during KMC. Dots represent individual participants and bars indicate group means. ***$p < .001$.

The quantified metrics of inter-brain network showed that both the inter-brain density

(IBD, delta: from 0.004 to 0.150; theta: from 0.051 to 0.081; alpha: from 0.030 to 0.162; and beta: from 0.009 to 0.034) and inter-brain strength (IBS, delta: 0.404 ± 0.003 vs. 0.459 ± 0.003, $t(57) = -14.4$, $p < .001$, Cohen's d = -1.891; theta: 0.261 ± 0.001 vs. 0.269 ± 0.001, $t(57) = -6.63$, $p < .001$, Cohen's d = -0.871; alpha 0.257 ± 0.001 vs. 0.275 ± 0.001, $t(57) = -14.51$, $p < .001$, Cohen's d = -1.905; beta 0.250 ± 0.001 vs. 0.266 ± 0.002, $t(57) = -6.43$, $p < .001$, Cohen's d = -0.844) increased during KMC **(Fig. 3B, D, F, H)**.

**The relationship between intra-brain and inter-brain measurements.**

To examine the relationship between intra- and inter-brain neural dynamics, we correlated the before-to-during KMC changes in inter-brain strength (ΔIBS) with the corresponding changes in infant intra-brain metrics (spectral power and network measures) within each frequency band. The results revealed positive correlations in the alpha band, where IBS was positively correlated with infant local efficiency ($r = .329$, 95% CI: [.078, .542], $p = .012$) and clustering coefficient ($r = .352$, 95% CI: [.103, .559], $p = .007$). Furthermore, in the beta band, IBS was positively correlated with infant small worldness ($r = .312$, 95% CI: [.058, .528], $p = .017$) **(Fig. 4)**. No significant correlations were found between IBS and intra-brain metrics in the other frequency bands.

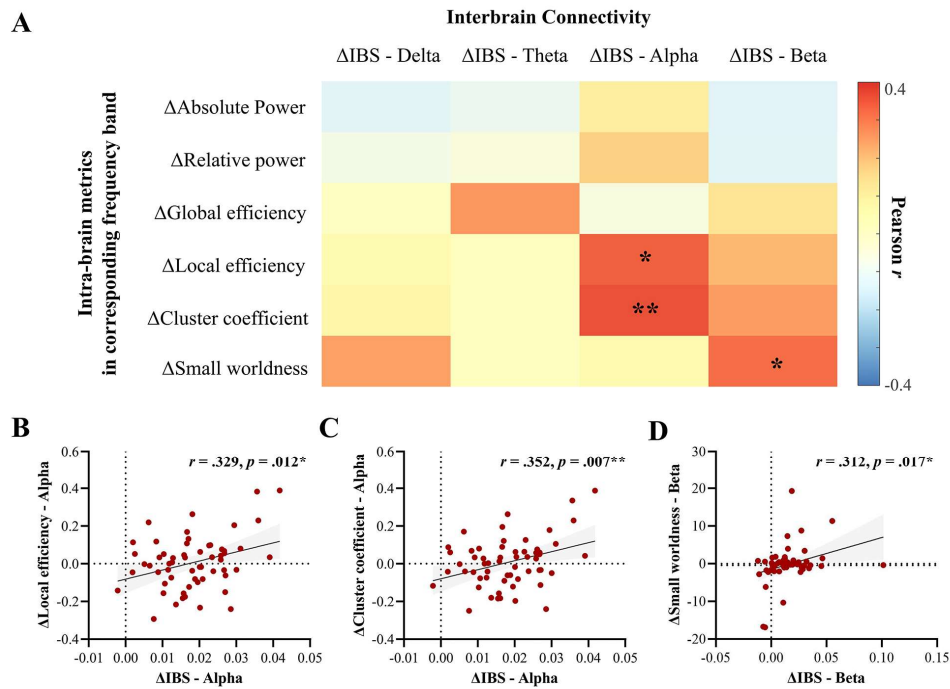

**Fig 4. Correlation between intra-brain and inter-brain measurements during KMC.** (A) HeatmCap showing correlations between the before-to-during KMC changes in inter-brain strength (IBS) and infant intra-brain metrics (absolute power, relative power, global efficiency, local efficiency, clustering coefficient, small-worldness) across frequency bands. *$p < .05$, **$p < .01$. (B) Scatter plot depicting the positive correlation between IBS and infant local efficiency in the alpha band. (C) Scatter plot depicting the positive correlation between IBS and infant clustering coefficient in the alpha band. (D) Scatter plot depicting the positive correlation between IBS and infant small-worldness in the beta band. Shaded areas represent 95% confidence intervals.

## Discussion

This study elucidates the immediate neurophysiological mechanisms by which KMC promotes development in preterm infants. Using EEG hyperscanning, we found that the initial contact during the first KMC not only enhances the preterm infant's intra-brain spectral power but also significantly increases the strength and density of mother-infant inter-brain connectivity. Furthermore, specific topological features of the infant's intra-brain network were positively correlated with inter-brain strength, suggesting that inter-brain synchrony during KMC may serve as a key mechanism for promoting single-brain development.

The significant change in EEG power during KMC, characterized by enhanced theta, alpha, and beta power alongside a reduced relative delta power, reflects an energy shift from lower to higher frequencies. Previous findings have demonstrated that, with advancing post-menstrual age, the EEG power spectrum of preterm infants shows a significant shift from lower to higher frequencies, characterized by decreased relative delta power and increased relative alpha and beta power [17,18]. Therefore, our findings implied KMC promoted infant's brain shifting toward a more mature spectral profile.

However, no statistically significant changes were observed in functional network topology metrics, including global efficiency, local efficiency, clustering coefficient and small-worldness. These results indicated that within the relatively short window of the first KMC, the infant brain's response is primarily characterized by enhanced intensity of electrocortical activity rather than a reconfiguration of network connectivity patterns. Given that the brain network topology mainly represents the architectural pattern of global information integration and is relatively more stable[19], our findings implied that the reorganization of brain networks may require repeated, sustained intervention to manifest.

This study is the first to use EEG hyperscanning to demonstrate that even during the initial contact between a preterm infant and mother, KMC significantly enhances inter-brain synchrony, suggesting that inter-brain synchrony is an immediate neurophysiological mechanism underlying KMC. This increase in inter-brain synchrony was observed across all frequency bands, particularly in the delta and alpha bands. Notably, the late-gestation fetal brain undergoes rapid development, with EEG changes primarily characterized by the amplitude and frequency of delta waves[20]; while the alpha rhythm is involved in cortical inhibition and sensory gating[21] and also plays a role in coordinating attention and expectation during interpersonal interaction[22] . Thus, the marked enhancement of mother-infant inter-brain synchrony in those bands suggested that KMC may facilitate both basic neurophysiological maturation and higher-order interactive processing. Furthermore, extending previous findings that parent-infant behavioral synchrony is not significantly evident in early infancy[15] , the analysis of inter-brain synchrony in this study reveals that KMC enhances mother-infant neural coupling, potentially offering an earlier and potentially more sensitive indicator of mother-infant attachment than behavioral synchrony alone.

Although infant single-brain network metrics did not show significant changes during KMC, those in alpha and beta bands were significantly correlated with mother-infant inter-brain strength. The results indicate that the increasing of mother-infant neural coupling during KMC is intrinsically linked to the infant's neurophysiological response. Specifically, infants who showed greater enhancement in inter-brain synchrony also exhibited more pronounced shifts toward a more optimized network efficiency. We therefore hypothesize that parent-infant inter-brain synchrony may be a crucial pathway through which KMC promotes long-term social function, empathy, and attachment in preterm infants. The level of parent-infant inter-brain synchrony could even be developed as a real-time marker for evaluating early emotional interaction and predicting long-term neurodevelopmental outcomes.

While this study innovatively proposes that inter-brain synchrony and intra-inter-brain interactions may represent a core neurophysiological mechanism of KMC, it has limitations. As a single-center study with a relatively limited sample size, the generalizability of the conclusions requires validation in larger, multi-center studies. Furthermore, EEG data were collected only during the first KMC session. Future longitudinal studies should track the dynamic evolution of single-brain function, inter-brain synchrony, and their interaction following cumulative effects of multiple KMC sessions. Correlating these measures with long-term neurodevelopmental outcomes (e.g., cognition, language, social function) would help establish a complete evidence chain from immediate neurophysiological effects to cumulative neurophysiological effects to long-term neurodevelopmental outcomes.

KMC is a simple, effective, and evidence-based intervention strongly recommended by the World Health Organization for global promotion, particularly in developing countries. However, its clinical implementation often remains "fragmented, intermittent, and far from sufficient," with healthcare providers frequently lacking confidence and experience in sustaining skin-to-skin contact during respiratory support or invasive procedures[5,6]. This study provides new neurophysiological evidence supporting the broader implementation of KMC. Our findings suggest that the neurophysiological mechanism of KMC extends beyond merely improving single-brain function; by enhancing mother-infant inter-brain neural synchrony, KMC may also promote intra-brain development in preterm infants, thereby establishing a neural foundation for their socio-emotional development and parent-infant attachment. Future research should explore whether the level of inter-brain synchrony during the first KMC session can predict long-term parent-infant behavioral synchrony and child neurodevelopmental outcomes. Ultimately, inter-brain synchrony metrics could be developed as objective neural markers for the early identification of at-risk dyads and the real-time evaluation of intervention effectiveness.

## Acknowledgements

We thank all premature infants, their parents and nurses who participated in the study.

## Fundings

This work was supported by the National Natural Science Foundation of China (No. 62371285), National Science and Technology Innovation 2030 Major Project of China (STI2030-Major Projects+2021ZD0204200), National Natural Science Foundation of China (82401803,82571697), Shanghai Pujiang Program(22PJC074).